\documentclass[sigconf]{acmart}

\AtBeginDocument{%
  \providecommand\BibTeX{{%
    \normalfont B\kern-0.5em{\scshape i\kern-0.25em b}\kern-0.8em\TeX}}}

\copyrightyear{2020}
\acmYear{2020}
\setcopyright{acmcopyright}\acmConference[LSC '20]{Proceedings of the Third Annual Workshop on the Lifelog Search Challenge}{June 9, 2020}{Dublin, Ireland}
\acmBooktitle{Proceedings of the Third Annual Workshop on the Lifelog Search Challenge (LSC '20), June 9, 2020, Dublin, Ireland}
\acmPrice{15.00}
\acmDOI{10.1145/3379172.3391721}
\acmISBN{978-1-4503-7136-0/20/06}

\begin{document}

\title{lifeXplore at the Lifelog Search Challenge 2020}
\renewcommand{\shorttitle}{lifeXplore at the Lifelog Search Challenge 2020}

\author{Andreas Leibetseder, \mbox{Klaus Schoeffmann}}
\affiliation{%
  \institution{Institute of Information Technology}
  \institution{Alpen-Adria University, 9020 Klagenfurt Austria}
}
\email{{aleibets|ks}@itec.aau.at}


\renewcommand{\shortauthors}{Leibetseder et al.}


%
%




\fancyhead{}

\begin{abstract}







Since its first iteration in 2018, the Lifelog Search Challenge (LSC) -- an interactive competition for retrieving lifelogging moments -- is co-located at the annual ACM International Conference on Multimedia Retrieval (ICMR) and has drawn international attention. With the goal of making an ever growing public lifelogging dataset searchable, several teams develop systems for quickly solving time-limited queries during the challenge. Having participated in both previous LSC iterations, i.e. LSC2018 and LSC2019, we present our lifeXplore system -- a video exploration and retrieval tool combining feature map browsing, concept search and filtering as well as hand-drawn sketching. The system is improved by including additional deep concept YOLO9000, optical character recognition (OCR) as well as adding uniform sampling as an alternative to the system's traditional underlying shot segmentation.


\end{abstract}

\maketitle

\section{Introduction} 

With the quantified self movement~\cite{lupton2016quantified}, self-tracking practices have gained widespread popularity. Followers of the movement hold the desire to gain and exhibit knowledge about themselves through numbers, which predominantly is linked to sportive activities. Nevertheless, there is ambition to go one step further and integrate collecting personal data into most aspects of one's daily life. So called lifeloggers~\cite{gurrin2014lifelogging} record and maintain a digital diary of their lives, at various levels of detail -- be it by automatically and periodically taking pictures of their every step, recording their heart rate throughout the day or even manually logging their calorie intake. Of course much more common activities certainly can also be attributed to the lifelogging movement. For instance, when uploading pictures of precious moments to social networks, often not only time and date of corresponding photos are included but even other attributes like their geographic location, rendering them very informative activity logs. Therefore, given that most content uploaded to platforms such as Facebook\footnote{\url{https://www.facebook.com}} are very likely to somehow be related to persons maintaining their own profiles, they at least in a basic sense can indeed be considered lifeloggers.

Although lifelogging can entail much personal gain such as the potential of recalling and reliving personal highlights, maintaining a life log can be costly: apart from certain lifestyle adjustments, technological advances as well as long-term storage have to be considered for the diverse and multimodal data collected. Finally, when considering the eventually overwhelming amount of data gathered over time, its maintenance and organization becomes harder and harder: how can a person revisit past memories without potentially knowing corresponding dates, places and other potentially important details? Albeit extensive lifelogs can certainly explored aimlessly, they can not easily be searched to find specific events in a way that complements human recollection typically fading over time. Therefore, it becomes apparent that managing big lifelogging archives requires smart interactive technologies not only for retrieval but also exploration, which possibly can trigger a person to remember more details.

\begin{figure*}[htbp!]
\center
  \includegraphics[width=\textwidth]{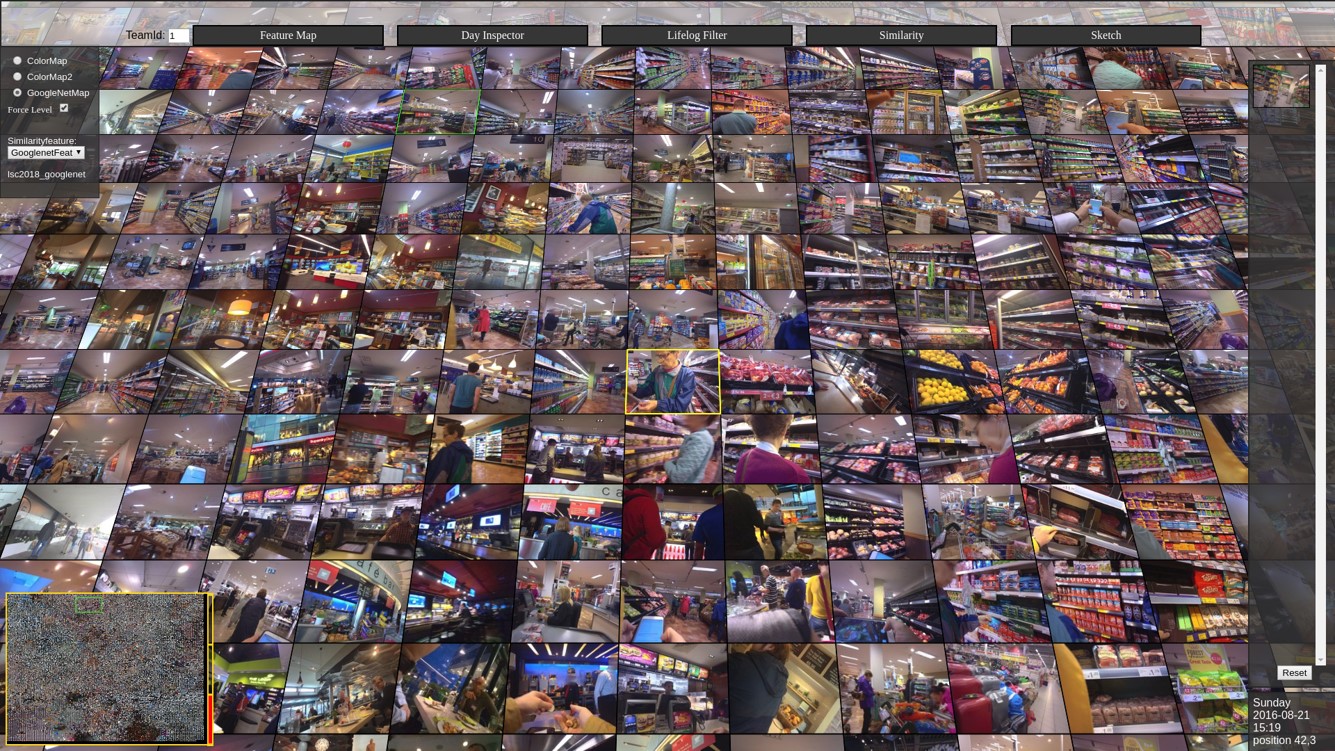}
  \caption{lifeXplore's feature maps: hierarchical 2D shot-keyframe grids organized according to various criteria.}
  \label{fig:interface}
\end{figure*}


The annual Lifelog Search Challenge~\cite{LSC20, gurrin2019comparing} (LSC), co-located at the ACM International Conference on Multimedia Retrieval (ICMR), addresses aforementioned issues and encourages international teams to develop systems for lifelog exploration and retrieval. It features a large public lifelogging dataset that needs to be made manageable by the participating teams. By adequately preprocessing and integrating the data into their individual systems prior to the competition, the teams later compete against each other in quickly retrieving specific lifelogging events, i.e. solving time-limited textual tasks announced at the conference during a live competition -- the fastest team finding the most events wins. Similar to the annual Video Browser Showdown~\cite{lokoc2019interactive,Lokoc2018,VBS2014} (VBS), co-located at the International Conference on MultiMedia Modeling (MMM), not only the systems' creators compete as \emph{expert} users but, in order to encourage the development of easy-to-use systems, uninvolved persons such as challenge spectators are as well asked to participate as \emph{novice} users, each handling one of the provided systems after being introduced to it briefly. In contrast to LSC2018 and LSC2019 employing an NTCIR-13 lifelogging test collection comprising roughly a month of collected data~\cite{gurrin2019test}, LSC2020's dataset~\cite{LSC20} is more than four times as big and contains several month-long periods of consecutive lifelogs recorded within the years of 2015-2018. Similar to the previous dataset, the one for LSC2020 consists of daily logs comprising photos taken periodically (roughly every 40 seconds) from a first-person perspective with automatically extracted deep visual concepts as well as time-stamped data from other sensors, e.g. geographic location, speed or heart rate.


As already employed for previous LSC iterations, for LSC2020 we as well propose our evolving lifeXplore system~\cite{leibetseder2019lifexplore,munzer2018lifexplore} -- see Figure~\ref{fig:interface} depicting its main interface. It is based on diveXplore~\cite{leibetseder2020divexplore, schoeffmann2019autopiloting, primus2018itec, leibetseder2018sketch, schoeffmann2017collaborative} -- a system iteratively developed for the most recent VBS challenges (VBS2017-VBS2020). Improvements to the current system are summarized as follows:

\begin{description}
    \item[YOLO9000:] We add YOLO9000 to our collection of deep concepts, cf. Section~\ref{YOLO}.
    \item[OCR:] We add optical character recognition (OCR) as a new filter, cf. Section~\ref{OCR}.
    \item[Uniform Sampling:] Additionally to our traditional shot segmentation, we add uniform sampling as an alternative way to segment videos, cf. Section~\ref{uniform_sampling}.
\end{description}


The remainder of this paper is structured as follows: we describe lifeXplore briefly in Section~\ref{lifexplore_description}, present example strategies of how to solve a past LSC task from an expert as well as novice perspective in Section~\ref{solving_lsc} and outline major changes and improvements for LSC2020 in Section~\ref{lifexplore_improvements}. Section~\ref{sec: conclusions} finally concludes this work together with giving an outlook on future work.

\section{The lifeXplore system}
\label{lifexplore_description}

The web-technologes-based lifeXplore system is derived from the interactive video exploration and retrieval system diveXplore, hence, it is build for working with pre-analyzed video clips. As the LSC dataset exclusively contains images, for preprocessing they need to be transcoded into videos: one video is created for a day's worth of sequential image data at a frame-rate of 5 fps -- this ensures an adequate playback rate for the in-system videoplayer and makes video navigation more precise than with faster frame-rates. This results in 114 video clips, which subsequently are organized into coherent scenes according to a custom shot detection algorithm~\cite{PrimusTrecVID2016} based on optical flow. For every shot detected, a keyframe is chosen that the system uses as a representation for its entire content on multiple views. For instance, Figure~\ref{fig:interface} shows the system's main interface -- a feature map view where all shot keyframes are ordered in a hierarchical two-dimensional grid according to different criteria~\cite{schoeffmann2017itec}, e.g. color. Finally, the keyframes are analyzed using deep models such as Inception-BN~\cite{ioffe2015batch} or custom descriptors like HistMap~\cite{schoeffmann2018howexperts}, extracting concepts as well as feature vectors for retrieval and similarity search.

\subsection{Exploration}

Multiple integrated strategies for dataset exploration offer a user powerful tools to get familiar with the dataset, which not only may prove useful during solving given tasks but also in-between them. In the following paragraphs, we describe said strategies in more detail.

\subsubsection{Feature Map Browsing}

The system's various feature maps offer a convenient way of exploring the underlying dataset according to different aspects. Albeit potentially too time-consuming for solving a task limited to a few minutes, an initial map exploration can yield a good starting position for further query refinement. Feature maps can be switched quickly and users can choose from a variety of maps organized by color, edge histogram, motion and individual concepts such as "car" or "person". As depicted in Figure~\ref{fig:interface}, a feature map typically contains much more keyframes than lifeXplore can display on one screen. Therefore, the lower left corner is used to display a toggleable minimap, which is used to indicate the current position within the entirety of keyframes as well as the map's current level -- users are able to zoom the map in and out in order to adjust the level of detail, i.e. the amount of keyframes, displayed. The number of map levels depends on the map's overall size -- Figure~\ref{fig:interface}'s feature map includes three levels, indicated by the three small bars at the minimap's right-hand side. Hovering the minimap gives the user a preview of the keyframe at the mouse position and it is possible to jump to corresponding location via a simple left-click. Additionally, maps can as well be automatically navigated by using the \emph{Autopilot} feature for scanning keyframes in a spiral pattern.

\begin{figure}[htbp!]
\center
\includegraphics[width=0.48\textwidth]{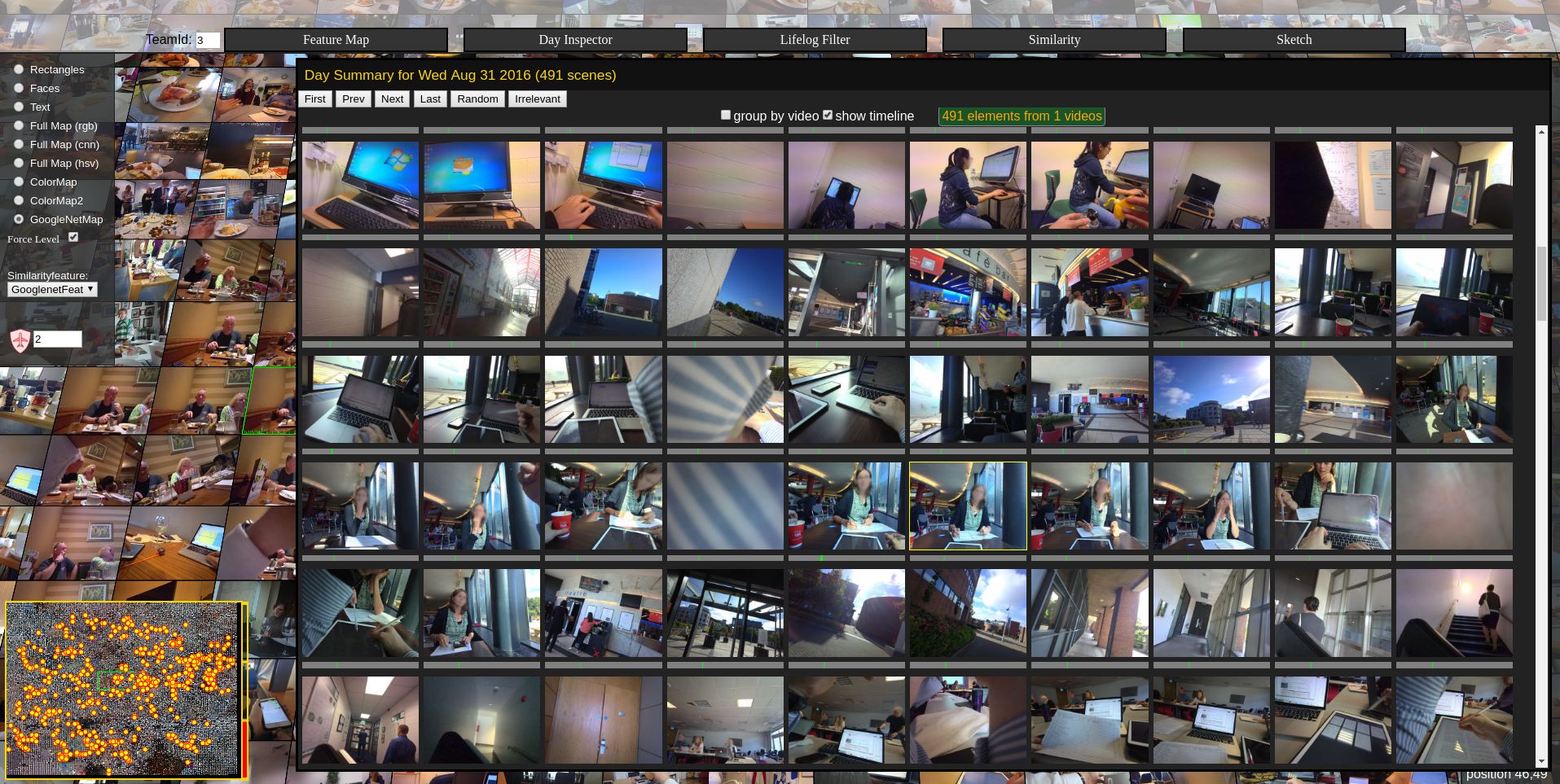}
\caption{Day summary view: contains all shots of a single day. Days can be switched via previous/next buttons as well as via their unique IDs.}
\label{fig:day_summary}
\end{figure}

\subsubsection{Day Summary View}

Additionally, lifeXplore offers a day summary view, shown in Figure~\ref{fig:day_summary}, that enables a quick overview of all shots of a specific day. This kind of exploration feature is even accessible from any view through a right-click menu, which can be opened for every keyframe. The summary view also facilitates browsing consecutive days by providing suitable navigation such as a "previous" and "next" button as well as a possibility for opening a specific day when entering its ID.

\begin{figure}[htbp!]
\center
\includegraphics[width=0.48\textwidth]{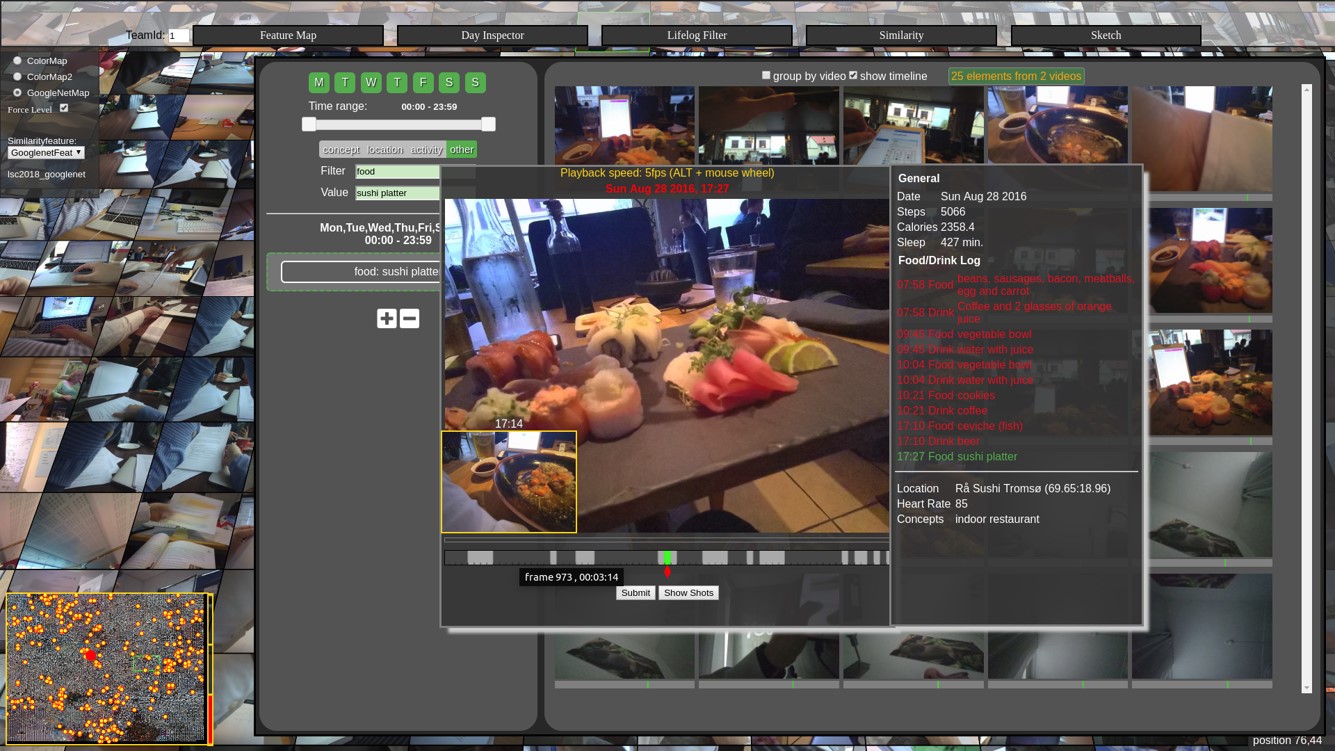}
\caption{Day inspector view: allows for browsing the video created from a day's worth of images. Also includes links to specific events defined by metadata.}
\label{fig:video_view}
\end{figure}

\subsubsection{Day Inspector View}

Finally, an integrated day inspector provides a third way of exploring the data, illustrated in Figure~\ref{fig:video_view}: by opening the underlying video and displaying shot-boundaries as well as -metadata, it enables even more detailed inspection than the day summary view. The view's video player, furthermore, offers standard controls and convenience functionalities like variable playback speed and the possibility for submitting the current frame in order to attempt solving the task at hand. 

\subsection{Retrieval}

There are several ways of retrieving content integrated into lifeXplore: concept and metadata search, sketch search as well as similarity search.

\begin{figure}[htbp!]
\center
\includegraphics[width=0.48\textwidth]{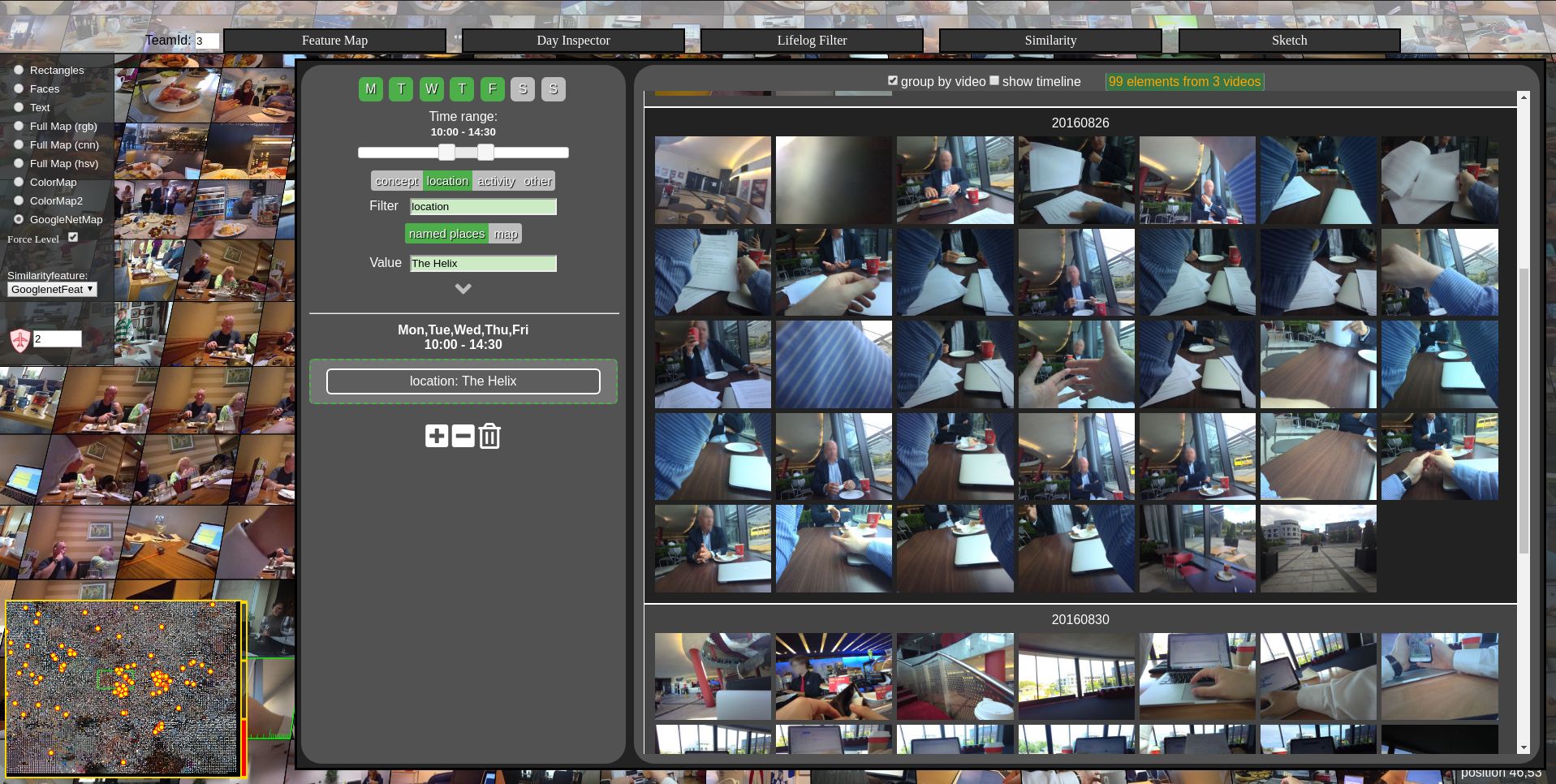}
\caption{Combinable drag and drop filtering: filters are applied via search or selection and combined using draggable containers.}
\label{fig:combinable_filters}
\end{figure}

\subsubsection{Filter View}

The filter view provides a user with a variety of combinable filters: as depicted in the overlay shown in Figure~\ref{fig:combinable_filters}, a user used the view's tools on the left-hand side to activate the weekday filter for only retrieving weekdays together with a time range filter for "10:00-14:30" in combination with a named location from the metadata, "The Helix". The result view on the right-hand side contains all retrieved shots ordered by video. Besides the most basic filters such as weekday and time, lifeXplore includes following additional filters: deep concepts, speed, heart rate, number of steps, calories burned, activity type and geographic location -- both, using metadata-provided named locations as well as latitude and longitude information, which is selectable via an offline map server\footnote{Openmaptiles: \url{https://openmaptiles.com/}}. All of these filters can be combined via specific filter containers that model the notion of logical AND and OR query concatenations. Altering said containers automatically issues all necessary queries and updates the displayed results on the right-hand side.

\begin{figure}[htbp!]
\center
\includegraphics[width=0.48\textwidth]{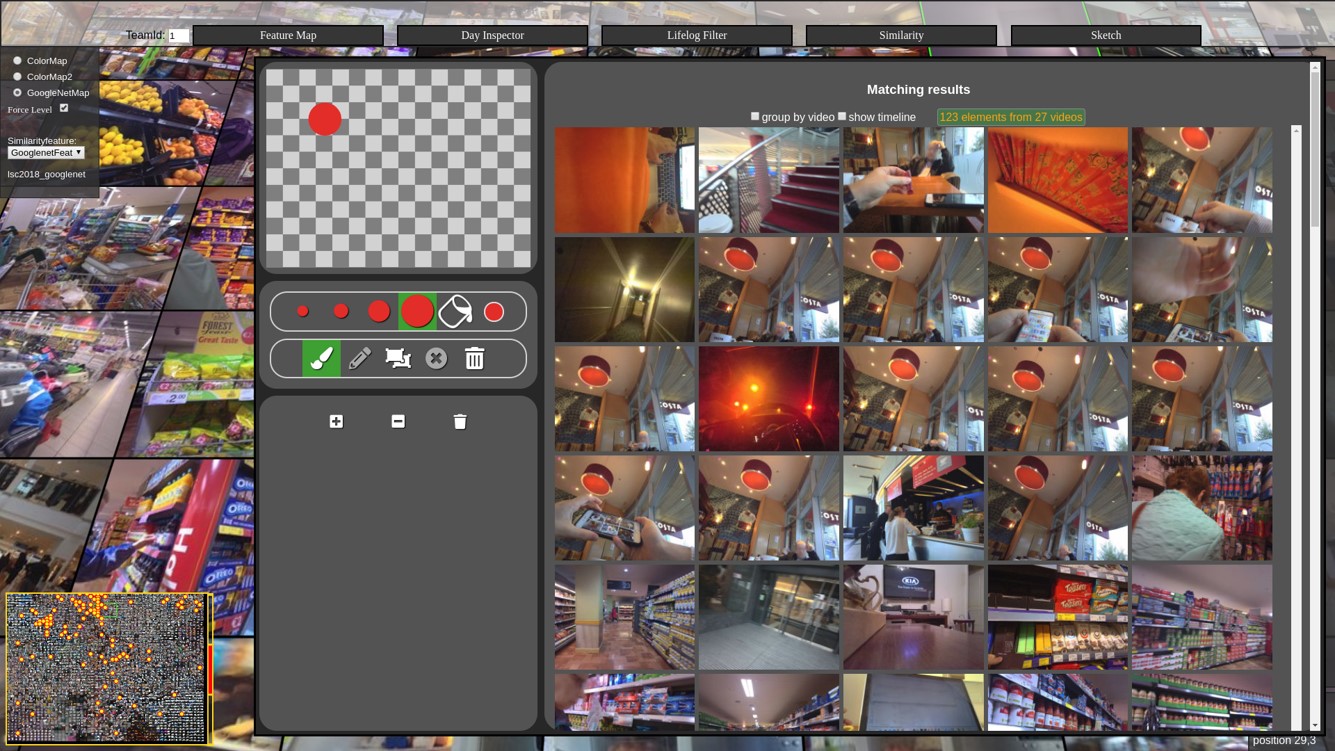}
\caption{Sketch search view: a reduced color palette and a drawing canvas are used to create simple sketches (e.g. a simple red dot). Matching shots are displayed on the right-hand side.}
\label{fig:sketch_search}
\end{figure}

\subsubsection{Sketch Search View}

The sketch search view, representing an alternate way of retrieving keyframes, offers a user to draw a simple sketch that is compared to the database using the custom developed HistMap descriptor~\cite{schoeffmann2018howexperts}. Figure~\ref{fig:sketch_search} shows this feature in action: after using a canvas placed on the top left corner of the overlay for drawing, HistMap is extracted and compared to the database with the result of displaying the top retrieved similar shots on the the right-hand side. As indicated in the figure, canvas areas left blank are ignored -- only the actually painted regions affect the result list.

\begin{figure}[htbp!]
\center
\includegraphics[width=0.48\textwidth]{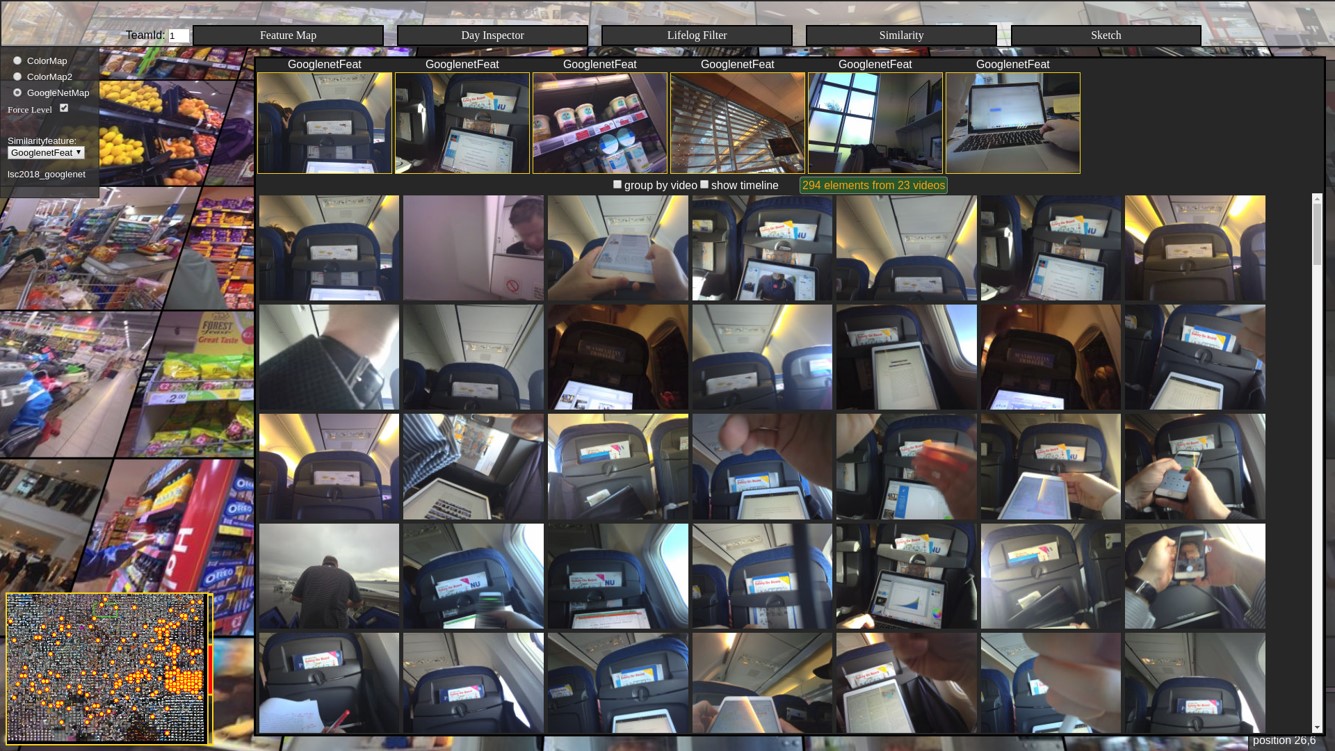}
\caption{Similarity search view: shows similar shots to user-chosen shots using selectable similarity metrics. Can be accessed from any of the system's shot views.}
\label{fig:similarity_search}
\end{figure}

\subsubsection{Similarity Search}

Lastly, the system also includes a similarity search, portrayed in Figure~\ref{fig:similarity_search}. Similarity is estimated according to deep network feature vectors as well as the HistMap descriptor: given any example keyframe, lifeXplore tries to find similar shots according to the current user-chosen similarity measure. In the figure, GoogLeNet features~\cite{szegedy2015going}, i.e. the feature vectors of the net's last fully connected layer, are used to quite successfully find similar shots to a back view of an airplane seat. Similarity search is as well accessible via the right-click menu opening for any shot keyframe.

\section{Solving a Lifelogging Task}
\label{solving_lsc}

In this section, we describe potential search strategies that novice as well as expert users might apply when solving following task from LSC2019:

\begin{itemize}
    \item[\bf t=0] Watching an important football game
    \item[\bf t=30] on a tablet computer.
    \item[\bf t=60] I had been drinking beer.
    \item[\bf t=90] Oh, I was at a table with friends when drinking beer,
    \item[\bf t=120] but returned to my hotel room to watch the football.
    \item[\bf t=150] I wasn’t using the TV in the room for some reason. I was using the tablet.
\end{itemize}

This textual search task was presented to the competing teams with an overall duration of 180 seconds and the timestamps (t) indicate at which second mark the information in any of above lines became available for the participants. The desired event is unique, meaning it only exists once in the whole dataset. Although the task originally was given as a novice task, we also include an expert's perspective below.

\subsection{Novice Strategies}

Since novice users typically need to be prepared for using a system very quickly, with lifeXplore they should most importantly be introduced to the filter view, where they can mix and match filters. Early on, i.e. in the first 60 seconds, they might create two concept filters with keywords such as "football", "sports" or "game" and "tablet" or "computer". At second 90 the search becomes a little trickier because lifeXplore does not yet include temporal connections between filters. Hence, either users can start a new filter search including new concepts, like "multiple people" and "beer", or ignore this information for the time being -- these circumstances, nevertheless, should be kept in mind, since they eventually are required to identify the correct scene. For example, when using the day summary for inspecting several videos including tablets that show football games, a user needs to investigate such prior events for determining the event in question. Finally with all the rest of the information available at seconds 120 and 150, it becomes clear that the lifelogger has been abroad during this event, therefore, a novice user may activate a location filter in order to ignore all locations marked as "home" as well as try out adding "TV" as an another concept filter. 

\subsection{Expert Strategies}

Experts have more experience with the systems as well as the dataset, therefore, in addition to the previously described novice strategies, they better know which concepts may yield acceptable results: e.g. the currently available dataset includes many shots that contain the concept "laptop", which oftentimes is detected instead of "tablet", thus, could alternatively be used for concept filtering. Also, their familiarity with sketch search could potentially incline them to try drawing a large green rectangle and combine it with backgrounds of various light intensities: a person watching a game on a tablet might sit very close to it and the grass field of a football match predominantly appears in green colors. Furthermore, finding green-shaded shots could as well be attempted by perusing a color-based feature map for locating clusters of green keyframes. Drawing from a larger pool of search experience, experts might further infer certain conditions during the task. For instance, being abroad might entail a longer period of travelling, which could be an extensive car ride or even a flight -- locating such events can at times greatly reduce the search space. As another example, drinking beer in a group is more likely to happen in the afternoon or evening, which might encourage a user to restrict the time of the day filter accordingly. Finally, having found some shots of tablets showing any content, an expert might turn to use similarity search with CNN features, which typically yields a good variety of results that may contain the correct tablet displaying the football game.

\section{Improvements}
\label{lifexplore_improvements}

In addition to constant interface and usability improvements, the system's major additions that are made for LSC2020 mostly include integrating new filter criteria and additional deep concepts. In this section, we deliberately omit smaller system improvements, as they often are not significant enough for the overall goal of the system and only focus on three selected enhancements: adding YOLO9000~\cite{redmon2017yolo9000} deep concepts, integrating OCR and including uniform sampling as an additional way of segmenting the day videos.

\subsection{Deep Concepts: YOLO9000}
\label{YOLO}

YOLO9000~\cite{redmon2017yolo9000} is a fast object detection system able to detect 9000+ object categories, which are hierarchically dependent on each other. Its integration into lifeXplore not only enables filtering using additional concepts but also provides bounding boxes of object locations, which are usable for potential future functionalities utilizing object detection: users could, for instance, assign concepts to their hand-drawn sketches allowing for the integration of an additional sketch-based strategy for locating objects.

\subsection{Optical Character Recognition (OCR)}
\label{OCR}

Since previous iterations of the LSC as well as VBS have shown the usefulness of text recognition, we as well include this functionality into lifeXplore. For extraction, we use the open source OCR engine \emph{Tesseract}\footnote{Tesseract:\url{https://github.com/tesseract-ocr}}. OCR is integrated as an additional combinable filter extending the functionalities of corresponding view and the system's database.

\subsection{Uniformly Sampled Segments}
\label{uniform_sampling}

As mentioned, the shot segmentation of lifeXplore is conducted using a custom designed algorithm using optical flow that targets traditional video content. Since the LSC data is image-based with large gaps between images, this approach sometimes exhibits the disadvantage of having too many or too few shots. To mitigate this problem we integrate uniform sampling as an alternative way of segmenting the day video clips. As a sampling rate we choose 10 frames, which, considering an approximate gap of 40 seconds between images, results in segments that include about 7 minutes of content. We implement this additional shot segmentation alongside the traditional one, as not to lose the advantages of coherent scene detection when functioning properly. This improvement triggers changes to all existing views, as the system was developed keeping shot exploration in mind. Nevertheless, the possibility of switching between multiple sampling rates greatly improves result exploration: e.g. when using the day summary view, quickly lowering the sample rate instantly shows a much more compact summary saving precious exploration time.

\section{Conclusions and Future Work} \label{sec: conclusions}

The annual Lifelog Search Challenge is an international competition for advancing in the field of interactive lifelog analysis and retrieval. Using custom developed systems, expert and novice users compete against each other to retrieve specific textually described scenes in a large dataset spanning close to 4 months of collected lifelogging data. With the goal of participating in LSC2020, we propose our improved \emph{lifeXplore} system, which already has been employed in previous challenge iterations. For future work, we are planning to address the continuous growth of lifelog data and further improve the scalability of our system.


\begin{acks}
This work was funded by the FWF Austrian Science Fund under grant P 32010-N38.
\end{acks}

\balance

%
\bibliographystyle{ACM-Reference-Format}
\bibliography{bibliography}


\begin{thebibliography}{21}


\ifx \showCODEN    \undefined \def \showCODEN     #1{\unskip}     \fi
\ifx \showDOI      \undefined \def \showDOI       #1{#1}\fi
\ifx \showISBNx    \undefined \def \showISBNx     #1{\unskip}     \fi
\ifx \showISBNxiii \undefined \def \showISBNxiii  #1{\unskip}     \fi
\ifx \showISSN     \undefined \def \showISSN      #1{\unskip}     \fi
\ifx \showLCCN     \undefined \def \showLCCN      #1{\unskip}     \fi
\ifx \shownote     \undefined \def \shownote      #1{#1}          \fi
\ifx \showarticletitle \undefined \def \showarticletitle #1{#1}   \fi
\ifx \showURL      \undefined \def \showURL       {\relax}        \fi
\providecommand\bibfield[2]{#2}
\providecommand\bibinfo[2]{#2}
\providecommand\natexlab[1]{#1}
\providecommand\showeprint[2][]{arXiv:#2}

\bibitem[\protect\citeauthoryear{Gurrin, Le, Ninh, Dang-Nguyen, Jónsson,
  Schoeffmann, Lokoč, Hurst, and Tran}{Gurrin et~al\mbox{.}}{2020}]%
        {LSC20}
\bibfield{author}{\bibinfo{person}{Cathal Gurrin}, \bibinfo{person}{Tu-Khiem
  Le}, \bibinfo{person}{Van-Tu Ninh}, \bibinfo{person}{Duc-Tien Dang-Nguyen},
  \bibinfo{person}{Björn~\THór Jónsson}, \bibinfo{person}{Klaus
  Schoeffmann}, \bibinfo{person}{Jakub Lokoč}, \bibinfo{person}{Wolfgang
  Hurst}, {and} \bibinfo{person}{Minh-Triet Tran}.}
  \bibinfo{year}{2020}\natexlab{}.
\newblock \showarticletitle{{An Introduction to the Third Annual Lifelog Search
  Challenge}}. In \bibinfo{booktitle}{\emph{ICMR '20, The 2020 International
  Conference on Multimedia Retrieval}}. \bibinfo{publisher}{ACM},
  \bibinfo{address}{Dublin, Ireland}.
\newblock


\bibitem[\protect\citeauthoryear{Gurrin, Schoeffmann, Joho, Leibetseder, Zhou,
  Duane, Nguyen, Tien, Riegler, Piras, et~al\mbox{.}}{Gurrin
  et~al\mbox{.}}{2019a}]%
        {gurrin2019comparing}
\bibfield{author}{\bibinfo{person}{Cathal Gurrin}, \bibinfo{person}{Klaus
  Schoeffmann}, \bibinfo{person}{Hideo Joho}, \bibinfo{person}{Andreas
  Leibetseder}, \bibinfo{person}{Liting Zhou}, \bibinfo{person}{Aaron Duane},
  \bibinfo{person}{Dang Nguyen}, \bibinfo{person}{Duc Tien},
  \bibinfo{person}{Michael Riegler}, \bibinfo{person}{Luca Piras},
  {et~al\mbox{.}}} \bibinfo{year}{2019}\natexlab{a}.
\newblock \showarticletitle{Comparing approaches to interactive lifelog search
  at the lifelog search challenge (LSC2018)}.
\newblock \bibinfo{journal}{\emph{ITE Transactions on Media Technology and
  Applications}} \bibinfo{volume}{7}, \bibinfo{number}{2}
  (\bibinfo{year}{2019}), \bibinfo{pages}{46--59}.
\newblock


\bibitem[\protect\citeauthoryear{Gurrin, Schoeffmann, Joho, Munzer, Albatal,
  Hopfgartner, Zhou, and Dang-Nguyen}{Gurrin et~al\mbox{.}}{2019b}]%
        {gurrin2019test}
\bibfield{author}{\bibinfo{person}{Cathal Gurrin}, \bibinfo{person}{Klaus
  Schoeffmann}, \bibinfo{person}{Hideo Joho}, \bibinfo{person}{Bernd Munzer},
  \bibinfo{person}{Rami Albatal}, \bibinfo{person}{Frank Hopfgartner},
  \bibinfo{person}{Liting Zhou}, {and} \bibinfo{person}{Duc-Tien Dang-Nguyen}.}
  \bibinfo{year}{2019}\natexlab{b}.
\newblock \showarticletitle{A test collection for interactive lifelog
  retrieval}. In \bibinfo{booktitle}{\emph{International Conference on
  Multimedia Modeling}}. Springer, \bibinfo{publisher}{Springer},
  \bibinfo{address}{Cham}, \bibinfo{pages}{312--324}.
\newblock


\bibitem[\protect\citeauthoryear{Gurrin, Smeaton, Doherty,
  et~al\mbox{.}}{Gurrin et~al\mbox{.}}{2014}]%
        {gurrin2014lifelogging}
\bibfield{author}{\bibinfo{person}{Cathal Gurrin}, \bibinfo{person}{Alan~F
  Smeaton}, \bibinfo{person}{Aiden~R Doherty}, {et~al\mbox{.}}}
  \bibinfo{year}{2014}\natexlab{}.
\newblock \showarticletitle{Lifelogging: Personal big data}.
\newblock \bibinfo{journal}{\emph{Foundations and Trends{\textregistered} in
  Information Retrieval}} \bibinfo{volume}{8}, \bibinfo{number}{1}
  (\bibinfo{year}{2014}), \bibinfo{pages}{1--125}.
\newblock


\bibitem[\protect\citeauthoryear{Ioffe and Szegedy}{Ioffe and Szegedy}{2015}]%
        {ioffe2015batch}
\bibfield{author}{\bibinfo{person}{Sergey Ioffe} {and}
  \bibinfo{person}{Christian Szegedy}.} \bibinfo{year}{2015}\natexlab{}.
\newblock \showarticletitle{Batch normalization: Accelerating deep network
  training by reducing internal covariate shift}. In
  \bibinfo{booktitle}{\emph{International conference on machine learning}}.
  \bibinfo{pages}{448--456}.
\newblock


\bibitem[\protect\citeauthoryear{Leibetseder, Kletz, and
  Schoeffmann}{Leibetseder et~al\mbox{.}}{2018}]%
        {leibetseder2018sketch}
\bibfield{author}{\bibinfo{person}{Andreas Leibetseder},
  \bibinfo{person}{Sabrina Kletz}, {and} \bibinfo{person}{Klaus Schoeffmann}.}
  \bibinfo{year}{2018}\natexlab{}.
\newblock \showarticletitle{Sketch-Based Similarity Search for Collaborative
  Feature Maps}. In \bibinfo{booktitle}{\emph{International Conference on
  Multimedia Modeling}}. Springer, \bibinfo{pages}{425--430}.
\newblock


\bibitem[\protect\citeauthoryear{Leibetseder, M{\"u}nzer, Primus, Kletz, and
  Schoeffmann}{Leibetseder et~al\mbox{.}}{2020}]%
        {leibetseder2020divexplore}
\bibfield{author}{\bibinfo{person}{Andreas Leibetseder}, \bibinfo{person}{Bernd
  M{\"u}nzer}, \bibinfo{person}{J{\"u}rgen Primus}, \bibinfo{person}{Sabrina
  Kletz}, {and} \bibinfo{person}{Klaus Schoeffmann}.}
  \bibinfo{year}{2020}\natexlab{}.
\newblock \showarticletitle{diveXplore 4.0: The ITEC Deep Interactive Video
  Exploration System at VBS2020}. In \bibinfo{booktitle}{\emph{International
  Conference on Multimedia Modeling}}. Springer, \bibinfo{pages}{753--759}.
\newblock


\bibitem[\protect\citeauthoryear{Leibetseder, M{\"u}nzer, Primus, Kletz,
  Schoeffmann, Berns, and Beecks}{Leibetseder et~al\mbox{.}}{2019}]%
        {leibetseder2019lifexplore}
\bibfield{author}{\bibinfo{person}{Andreas Leibetseder}, \bibinfo{person}{Bernd
  M{\"u}nzer}, \bibinfo{person}{Manfred~J{\"u}rgen Primus},
  \bibinfo{person}{Sabrina Kletz}, \bibinfo{person}{Klaus Schoeffmann},
  \bibinfo{person}{Fabian Berns}, {and} \bibinfo{person}{Christian Beecks}.}
  \bibinfo{year}{2019}\natexlab{}.
\newblock \showarticletitle{lifeXplore at the Lifelog Search Challenge 2019}.
  In \bibinfo{booktitle}{\emph{Proc. of the ACM Workshop on Lifelog Search
  Challenge}}. ACM, \bibinfo{pages}{13--17}.
\newblock


\bibitem[\protect\citeauthoryear{Lokoc, Bailer, Schoeffmann, Muenzer, and
  Awad}{Lokoc et~al\mbox{.}}{2018}]%
        {Lokoc2018}
\bibfield{author}{\bibinfo{person}{J. Lokoc}, \bibinfo{person}{W. Bailer},
  \bibinfo{person}{K. Schoeffmann}, \bibinfo{person}{B. Muenzer}, {and}
  \bibinfo{person}{G. Awad}.} \bibinfo{year}{2018}\natexlab{}.
\newblock \showarticletitle{On influential trends in interactive video
  retrieval: Video Browser Showdown 2015-2017}.
\newblock \bibinfo{journal}{\emph{IEEE Transactions on Multimedia}}
  (\bibinfo{year}{2018}), \bibinfo{pages}{1--1}.
\newblock
\showISSN{1520-9210}
\urldef\tempurl%
\url{https://doi.org/10.1109/TMM.2018.2830110}
\showDOI{\tempurl}


\bibitem[\protect\citeauthoryear{Loko\v{c}, Koval\v{c}\'{\i}k, M\"{u}nzer,
  Sch\"{o}ffmann, Bailer, Gasser, Vrochidis, Nguyen, Rujikietgumjorn, and
  Barthel}{Loko\v{c} et~al\mbox{.}}{2019}]%
        {lokoc2019interactive}
\bibfield{author}{\bibinfo{person}{Jakub Loko\v{c}}, \bibinfo{person}{Gregor
  Koval\v{c}\'{\i}k}, \bibinfo{person}{Bernd M\"{u}nzer},
  \bibinfo{person}{Klaus Sch\"{o}ffmann}, \bibinfo{person}{Werner Bailer},
  \bibinfo{person}{Ralph Gasser}, \bibinfo{person}{Stefanos Vrochidis},
  \bibinfo{person}{Phuong~Anh Nguyen}, \bibinfo{person}{Sitapa
  Rujikietgumjorn}, {and} \bibinfo{person}{Kai~Uwe Barthel}.}
  \bibinfo{year}{2019}\natexlab{}.
\newblock \showarticletitle{Interactive Search or Sequential Browsing? A
  Detailed Analysis of the Video Browser Showdown 2018}.
\newblock \bibinfo{journal}{\emph{ACM Trans. Multimedia Comput. Commun. Appl.}}
  \bibinfo{volume}{15}, \bibinfo{number}{1}, Article \bibinfo{articleno}{29}
  (\bibinfo{date}{Feb.} \bibinfo{year}{2019}), \bibinfo{numpages}{18}~pages.
\newblock
\showISSN{1551-6857}
\urldef\tempurl%
\url{https://doi.org/10.1145/3295663}
\showDOI{\tempurl}


\bibitem[\protect\citeauthoryear{Lupton}{Lupton}{2016}]%
        {lupton2016quantified}
\bibfield{author}{\bibinfo{person}{Deborah Lupton}.}
  \bibinfo{year}{2016}\natexlab{}.
\newblock \bibinfo{booktitle}{\emph{The quantified self}}.
\newblock \bibinfo{publisher}{John Wiley \& Sons}.
\newblock


\bibitem[\protect\citeauthoryear{M\"{u}nzer, Leibetseder, Kletz, Primus, and
  Schoeffmann}{M\"{u}nzer et~al\mbox{.}}{2018}]%
        {munzer2018lifexplore}
\bibfield{author}{\bibinfo{person}{Bernd M\"{u}nzer}, \bibinfo{person}{Andreas
  Leibetseder}, \bibinfo{person}{Sabrina Kletz},
  \bibinfo{person}{Manfred~J\"{u}rgen Primus}, {and} \bibinfo{person}{Klaus
  Schoeffmann}.} \bibinfo{year}{2018}\natexlab{}.
\newblock \showarticletitle{lifeXplore at the Lifelog Search Challenge 2018}.
  In \bibinfo{booktitle}{\emph{Proceedings of the 2018 ACM Workshop on The
  Lifelog Search Challenge}} \emph{(\bibinfo{series}{LSC '18})}.
  \bibinfo{publisher}{ACM}, \bibinfo{address}{New York, NY, USA},
  \bibinfo{pages}{3--8}.
\newblock
\showISBNx{978-1-4503-5796-8}
\urldef\tempurl%
\url{https://doi.org/10.1145/3210539.3210541}
\showDOI{\tempurl}


\bibitem[\protect\citeauthoryear{Primus, M{\"u}nzer, Leibetseder, and
  Schoeffmann}{Primus et~al\mbox{.}}{2018}]%
        {primus2018itec}
\bibfield{author}{\bibinfo{person}{Manfred~J{\"u}rgen Primus},
  \bibinfo{person}{Bernd M{\"u}nzer}, \bibinfo{person}{Andreas Leibetseder},
  {and} \bibinfo{person}{Klaus Schoeffmann}.} \bibinfo{year}{2018}\natexlab{}.
\newblock \showarticletitle{The ITEC Collaborative Video Search System at the
  Video Browser Showdown 2018}. In \bibinfo{booktitle}{\emph{MultiMedia
  Modeling}}, \bibfield{editor}{\bibinfo{person}{Klaus Schoeffmann},
  \bibinfo{person}{Thanarat~H. Chalidabhongse}, \bibinfo{person}{Chong~Wah
  Ngo}, \bibinfo{person}{Supavadee Aramvith}, \bibinfo{person}{Noel~E.
  O'Connor}, \bibinfo{person}{Yo-Sung Ho}, \bibinfo{person}{Moncef Gabbouj},
  {and} \bibinfo{person}{Ahmed Elgammal}} (Eds.). \bibinfo{publisher}{Springer
  International Publishing}, \bibinfo{address}{Cham},
  \bibinfo{pages}{438--443}.
\newblock
\showISBNx{978-3-319-73600-6}


\bibitem[\protect\citeauthoryear{Primus, M\"{u}nzer, Petscharnig, and
  Schoeffmann}{Primus et~al\mbox{.}}{2016}]%
        {PrimusTrecVID2016}
\bibfield{author}{\bibinfo{person}{Manfred~J\"{u}rgen Primus},
  \bibinfo{person}{Bernd M\"{u}nzer}, \bibinfo{person}{Stefan Petscharnig},
  {and} \bibinfo{person}{Klaus Schoeffmann}.} \bibinfo{year}{2016}\natexlab{}.
\newblock \showarticletitle{ITEC-UNIKLU Ad-Hoc Video Search Submission 2016}.
  In \bibinfo{booktitle}{\emph{Proceedings of TRECVID 2016}},
  \bibfield{editor}{\bibinfo{person}{George Awad}, \bibinfo{person}{Jonathan
  Fiscus}, \bibinfo{person}{Martial Michel}, \bibinfo{person}{David Joy},
  \bibinfo{person}{Wessel Kraaij}, \bibinfo{person}{Alan~F Smeaton},
  \bibinfo{person}{Georges Quénot}, \bibinfo{person}{Maria Eskevich},
  \bibinfo{person}{Robin Aly}, \bibinfo{person}{Gareth J~F Jones},
  \bibinfo{person}{Roeland Ordelman}, \bibinfo{person}{Benoit Huet}, {and}
  \bibinfo{person}{Martha Larson}} (Eds.). \bibinfo{publisher}{NIST, USA},
  \bibinfo{address}{NIST, Gaithersburg, MD, USA}, \bibinfo{pages}{10}.
\newblock


\bibitem[\protect\citeauthoryear{Redmon and Farhadi}{Redmon and
  Farhadi}{2017}]%
        {redmon2017yolo9000}
\bibfield{author}{\bibinfo{person}{Joseph Redmon} {and} \bibinfo{person}{Ali
  Farhadi}.} \bibinfo{year}{2017}\natexlab{}.
\newblock \showarticletitle{YOLO9000: better, faster, stronger}. In
  \bibinfo{booktitle}{\emph{Proceedings of the IEEE conference on computer
  vision and pattern recognition}}. \bibinfo{pages}{7263--7271}.
\newblock


\bibitem[\protect\citeauthoryear{Schoeffmann}{Schoeffmann}{2014}]%
        {VBS2014}
\bibfield{author}{\bibinfo{person}{Klaus Schoeffmann}.}
  \bibinfo{year}{2014}\natexlab{}.
\newblock \showarticletitle{A User-Centric Media Retrieval Competition: The
  Video Browser Showdown 2012-2014}.
\newblock \bibinfo{journal}{\emph{MultiMedia, IEEE}} \bibinfo{volume}{21},
  \bibinfo{number}{4} (\bibinfo{date}{Oct} \bibinfo{year}{2014}),
  \bibinfo{pages}{8--13}.
\newblock
\showISSN{1070-986X}
\urldef\tempurl%
\url{https://doi.org/10.1109/MMUL.2014.56}
\showDOI{\tempurl}


\bibitem[\protect\citeauthoryear{Schoeffmann, M{\"u}nzer, Leibetseder, Primus,
  and Kletz}{Schoeffmann et~al\mbox{.}}{2019}]%
        {schoeffmann2019autopiloting}
\bibfield{author}{\bibinfo{person}{Klaus Schoeffmann}, \bibinfo{person}{Bernd
  M{\"u}nzer}, \bibinfo{person}{Andreas Leibetseder},
  \bibinfo{person}{J{\"u}rgen Primus}, {and} \bibinfo{person}{Sabrina Kletz}.}
  \bibinfo{year}{2019}\natexlab{}.
\newblock \showarticletitle{Autopiloting Feature Maps: The Deep Interactive
  Video Exploration (diveXplore) System at VBS2019}. In
  \bibinfo{booktitle}{\emph{Intl. Conference on Multimedia Modeling}}.
  Springer, \bibinfo{pages}{585--590}.
\newblock


\bibitem[\protect\citeauthoryear{Schoeffmann, M{\"{u}}nzer, Primus, Kletz, and
  Leibetseder}{Schoeffmann et~al\mbox{.}}{2018}]%
        {schoeffmann2018howexperts}
\bibfield{author}{\bibinfo{person}{Klaus Schoeffmann}, \bibinfo{person}{Bernd
  M{\"{u}}nzer}, \bibinfo{person}{Manfred~J{\"{u}}rgen Primus},
  \bibinfo{person}{Sabrina Kletz}, {and} \bibinfo{person}{Andreas
  Leibetseder}.} \bibinfo{year}{2018}\natexlab{}.
\newblock \showarticletitle{How Experts Search Different than Novices - An
  Evaluation of the Divexplore Video Retrieval System at Video Browser Showdown
  2018}. In \bibinfo{booktitle}{\emph{2018 {IEEE} International Conference on
  Multimedia {\&} Expo Workshops, {ICME} Workshops 2018, San Diego, CA, USA,
  July 23-27, 2018}}. \bibinfo{publisher}{{IEEE} Computer Society},
  \bibinfo{pages}{1--6}.
\newblock
\urldef\tempurl%
\url{https://doi.org/10.1109/ICMEW.2018.8551552}
\showDOI{\tempurl}


\bibitem[\protect\citeauthoryear{Schoeffmann, Primus, Muenzer, Petscharnig,
  Karisch, Xu, and Huerst}{Schoeffmann et~al\mbox{.}}{2017a}]%
        {schoeffmann2017collaborative}
\bibfield{author}{\bibinfo{person}{Klaus Schoeffmann},
  \bibinfo{person}{Manfred~J{\"u}rgen Primus}, \bibinfo{person}{Bernd Muenzer},
  \bibinfo{person}{Stefan Petscharnig}, \bibinfo{person}{Christof Karisch},
  \bibinfo{person}{Qing Xu}, {and} \bibinfo{person}{Wolfgang Huerst}.}
  \bibinfo{year}{2017}\natexlab{a}.
\newblock \showarticletitle{Collaborative feature maps for interactive video
  search}. In \bibinfo{booktitle}{\emph{Intl. Conference on Multimedia
  Modeling}}. Springer, \bibinfo{pages}{457--462}.
\newblock


\bibitem[\protect\citeauthoryear{Schoeffmann, Primus, Muenzer, Petscharnig,
  Karisch, Xu, and Huerst}{Schoeffmann et~al\mbox{.}}{2017b}]%
        {schoeffmann2017itec}
\bibfield{author}{\bibinfo{person}{Klaus Schoeffmann},
  \bibinfo{person}{Manfred~J{\"u}rgen Primus}, \bibinfo{person}{Bernd Muenzer},
  \bibinfo{person}{Stefan Petscharnig}, \bibinfo{person}{Christof Karisch},
  \bibinfo{person}{Qing Xu}, {and} \bibinfo{person}{Wolfgang Huerst}.}
  \bibinfo{year}{2017}\natexlab{b}.
\newblock \showarticletitle{Collaborative Feature Maps for Interactive Video
  Search}. In \bibinfo{booktitle}{\emph{MultiMedia Modeling}},
  \bibfield{editor}{\bibinfo{person}{Laurent Amsaleg},
  \bibinfo{person}{Gylfi~Þ{\'o}r Guðmundsson}, \bibinfo{person}{Cathal
  Gurrin}, \bibinfo{person}{Bj{\"o}rn~Þ{\'o}r J{\'o}nsson}, {and}
  \bibinfo{person}{Shin'ichi Satoh}} (Eds.). \bibinfo{publisher}{Springer
  International Publishing}, \bibinfo{address}{Cham},
  \bibinfo{pages}{457--462}.
\newblock
\showISBNx{978-3-319-51814-5}


\bibitem[\protect\citeauthoryear{Szegedy, Liu, Jia, Sermanet, Reed, Anguelov,
  Erhan, Vanhoucke, and Rabinovich}{Szegedy et~al\mbox{.}}{2015}]%
        {szegedy2015going}
\bibfield{author}{\bibinfo{person}{Christian Szegedy}, \bibinfo{person}{Wei
  Liu}, \bibinfo{person}{Yangqing Jia}, \bibinfo{person}{Pierre Sermanet},
  \bibinfo{person}{Scott Reed}, \bibinfo{person}{Dragomir Anguelov},
  \bibinfo{person}{Dumitru Erhan}, \bibinfo{person}{Vincent Vanhoucke}, {and}
  \bibinfo{person}{Andrew Rabinovich}.} \bibinfo{year}{2015}\natexlab{}.
\newblock \showarticletitle{Going deeper with convolutions}. In
  \bibinfo{booktitle}{\emph{Proc. of the IEEE conference on computer vision and
  pattern recognition}}. \bibinfo{pages}{1--9}.
\newblock


\end{thebibliography}

\end{document}